      [1999/12/01 v1.4c Il Nuovo Cimento]
\documentclass{cimento}

\usepackage{graphicx}  

\title{The prompt emission \& peculiar break of GRB\,060124}

\author{P.A.~Curran\from{ins:api}\ETC\thanks{pcurran@science.uva.nl},
D.A.~Kann\from{ins:taut},
P.~Ferrero\from{ins:taut},
E.~Rol\from{ins:leic},
R.A.M.J.~Wijers\from{ins:api}}

\instlist{
\inst{ins:api} Astronomical Institute, University of Amsterdam, Netherlands
  \inst{ins:taut} Th\"uringer Landessternwarte Tautenburg, Germany
    \inst{ins:leic} Department of Physics \& Astronomy, University of Leicester, UK
}

\PACSes{
\PACSit{98.70.Rz}{$\gamma$-ray sources; $\gamma$-ray bursts}
\PACSit{95.85.Kr}{Visible}% (390-750 nm)}
\PACSit{95.85.Nv}{X-ray}
}

\begin{document}

\maketitle

\begin{abstract}

Our multi-wavelength analysis of GRB\,060124 shows the unusual behaviour of the decaying lightcurve as well as supporting the recently proposed phenomenon of long-lasting central engine activity. 
The prompt X-ray emission displays uncommonly well resolved flaring behaviour, with spectral evolution -- indicative of central engine activity -- which allows us to estimate the energy injection time for each flare. The otherwise smooth X-ray \& optical afterglows demonstrate achromatic breaks at about 1 day which differ significantly from the usual jet break in the blastwave model of afterglows.

\end{abstract}

%------------------------------------------------------------------------

\section{Introduction}
On 2006 January 24 the bright gamma-ray burst, GRB\,060124, triggered instruments on multiple satellites, including Swift and  Konus-Wind. Swift immediately slewed to the burst and started observing in X-rays, $\sim350$s before the main period of activity. The discovery of the bright optical counterpart only 1 hour after the burst allowed for well-sampled follow-up.  The brightness of the burst permitted observations in X-rays for nearly a month despite a redshift of $z = 2.3$.

\section{Prompt Emission}

We analysed the previously observed high-energy flaring from the Swift BAT \& XRT instruments and Konus-Wind\cite{ref:kann},\cite{ref:romano}. This exhibits spectral evolution, supporting the theory of a central engine which is active for extended periods. 
Our joint spectral fit of the data (0.2keV -- 1160keV) gives an average spectral index of $\beta= 0.5$ after a break at 1.2keV and in combination with negligible Hydrogen column density above that of the Galactic value. This is in contrast to previous interpretations of the spectrum\cite{ref:romano}.

We also fit the XRT data to the curvature effect model\cite{ref:curve}, and estimate the times at which the central engine ejects the energy required to power the two main flares. Our results are consistent with similar fits to other GRBs\cite{ref:liang}, in that the ejection time of flares is during the rising period of the flare, though ours is a most striking example.

%------------------------------------------------------------------------

\section{Peculiar Break}

\begin{figure} 
  \centering 
  \includegraphics[angle=0,width=90mm]{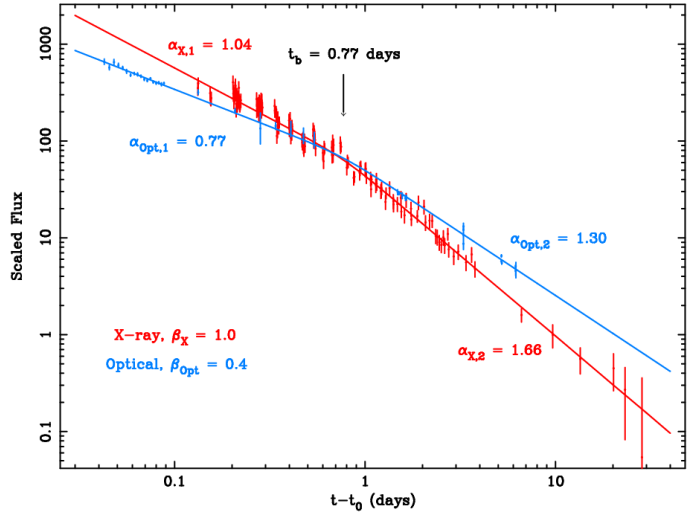} 
  %\caption{The X-ray (dashed) and optical (solid) lightcurves of GRB\,060124.} 
  \caption{The X-ray and optical lightcurves of GRB\,060124.} 
  \label{lc} 
\end{figure}

The fading optical \& X-ray lightcurves (\cite{ref:kann}; Fig. \ref{lc}) display simultaneous achromatic breaks. Before the break the temporal decay indices,  and spectral decay indices,  are consistent with the blastwave model of GRB afterglows in a homogeneous circumburst medium, with a power law index of the electron energy distribution, $p \approx 2$. The achromatic nature of the break points to a change in the dynamics of the blastwave. 

After the break there is a deviation from the behaviour expected from a laterally spreading jet. As observed in the majority of bursts, the temporal decay equals $p$, in both optical \& X-rays. Our analyses show shallower and unequal decay in both wavelength regimes. A possible explanation for this peculiar behaviour is that the jet is spreading at lower velocities than usually observed.

\end{document}